# A Comparative study of Analog and digital Controller On DC/DC Buck-Boost Converter Four Switch for Mobile Device Applications

Benlafkih Abdessamad[1], Krit Salah-ddine[2] and Chafik Elidrissi Mohamed[3]

[1]Ph.D student at Laboratory of physic and environment, Department of Physique, Faculty of Sciences, University Ibn Tofail, BP 133, Kénitra 14000, Morocco

[2]Asstt. Prof., Department of informatics, Polydisciplinary Faculty of Ouarzazate, University Ibn Zohr, BP/638, Morocco.

[3]Prof. Laboratory of physic and environment, Department of Physique, Faculty of Sciences, University Ibn Tofail, BP 133, Kénitra 14000, Morocco

a.benlafkih@yahoo.com, Krit_salah@yahoo.fr, and chafik_idrissi@yahoo.com

*Abstract*—this paper presents comparative performance between Analog and digital controller on DC/DC buck-boost converter four switch. The design of power electronic converter circuit with the use of closed loop scheme needs modeling and then simulating the converter using the modeled equations. This can easily be done with the help of state equations and MATLAB/SIMULINK as a tool for simulation of those state equations. DC/DC Buck-boost converter in this study is operated in buck (step-down) and boost (step-up) modes.

## I. INTRODUCTION

CURRENT trends in consumer electronics demand progressively lower supply voltages due to the unprecedented growth and use of wireless appliances. Portable devices, such as laptop computers and personal communication devices require ultra low-power circuitry to enable longer battery operation. The key to reducing power consumption while maintaining computational throughput and quality of service is to use such systems at the lowest possible supply voltage. The terminal voltage of the battery used in portable applications (e.g., NiMH, NiCd, and Li-ion) varies considerably depending on the state of their charging condition. For example, a single NiMH battery cell is fully charged to 1.8 V but it drops to 0.9 V before fully discharged [1]. Therefore, systems designed for a nominal supply voltage (say, 1.5 V with a single NiMH battery cell) require a converter capable of both stepping-up and stepping-down the battery voltage. While both buck (step-down) [2][3]and boost (step-up) [4]converters are widely used in power management circuits. The DC/DC Converter must provide a regulated DC output voltage even when varying load or the input voltage varies.

Therefore, the topologies for generating a voltage higher and lower than the supply is : non inverting buck-boost converter [5], [6],[7] which is essentially achieved by cascading a buck with a boost converter Fig.1, The trend in portable applications is to use the topologies that incorporate less number of external components and move closer to cost effective SOC designs [13].

Controller design for any system needs knowledge about system behavior. Usually this involves a mathematical description of the relation among inputs to the process, state variables, and output. This description in the form of mathematical equations which describe behavior of the system (process) is called model of the system [8][9]. This paper describes an efficient method to learn, analyze and simulation of DC/DC buck-boost converter four switch, with analog and digital Controller, The MATLAB/SIMULINK software package can be advantageously used to simulate power converters.

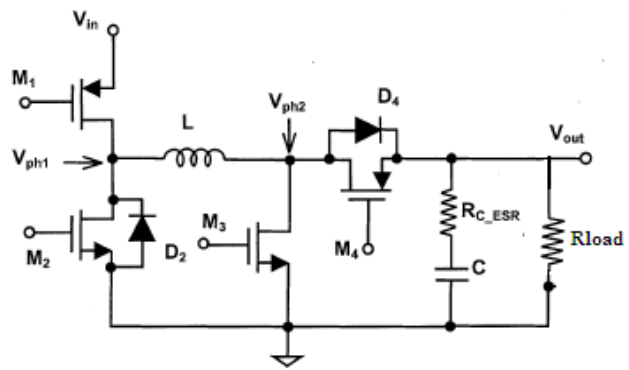

Fig.1 Noninverting synchronous buck-boost dc–dc converter.

## II. BUCK-BOOST CONVETER MODILING

*A. Open –loop synchonous buck-boost converter model*

In Fig.1 a DC-DC buck-boost converter is shown. The switching period is T and the duty cycle is D. Assuming continuous conduction mode of operation, During $T_{ON}$ the period of the cycle, switches $M_1$ and $M_3$ are ON and the input voltage is impressed across the inductor. Since the load current is instantaneously provided by the output capacitor during this interval, the capacitor voltage (output voltage) decreases, the state space equations are given by [9],

$$\begin{cases} \dfrac{di_L}{dt} = \dfrac{1}{L}\left[V_{in} - (R_L + R_{ON1} + R_{ON3}) \times i_L\right] \\ \dfrac{dv_c}{dt} = -\dfrac{1}{c} i_{out} \\ V_{out} = -R_{ESR} i_{out} + v_c \end{cases} \quad (1)$$

During the other interval of the switching period ($T_{OFF}$), switches $M_2$ and $M_4$ are turned ON and the inductor energy is transferred to the output, providing both the load current and also charging the output capacitor, the equations are given by [9],

$$\begin{cases} \dfrac{di_L}{dt} = -\dfrac{1}{L}[(R_L + R_{ON2} + R_{ON4}) \times i_L + V_{out}] \\ \dfrac{dv_c}{dt} = \dfrac{1}{c}(i_L - i_{out}) \\ V_{out} = -R_{ESR}i_{out} + v_c + R_{ESR}i_L \end{cases} \quad (2)$$

There is a time delay (known as dead-time) between turning OFF $M_1, M_3$ and turning ON $M_2, M_4$ to prevent shoot-through current. During this period, the inductor current flows through body diodes $D_2$ and $D_4$, from transistors $M_2$ and $M_4$, respectively.

The duty cycle (D) of the converter is given by
$$D = \frac{T_{ON}}{T_{ON}+T_{OFF}} = \frac{T_{ON}}{T} \quad (3)$$
Since the node $V_{ph1}$ is connected to $V_{in}$ for DT time over a period of T, the average voltage $V_{ph1,avg} = DV_{in}$. Similarly, the average node voltage of $V_{ph2}$ can be given by
$V_{ph2,avg} = D'V_{out}$ $(D' = 1 - D)$. Under steady-state operating condition, the inductor can be treated as short and the average voltage of $V_{ph1}$ and $V_{ph2}$ are equal
$$DV_{in} = D'V_{out} \implies \frac{V_{out}}{V_{in}} = \frac{D}{1-D} \quad (4)$$
The equations (1) and (2) are implemented in Simulink as shown in Fig. 2 to obtain the states, $i_L(t)$ and $V_{out}(t)$ [10][11][12].

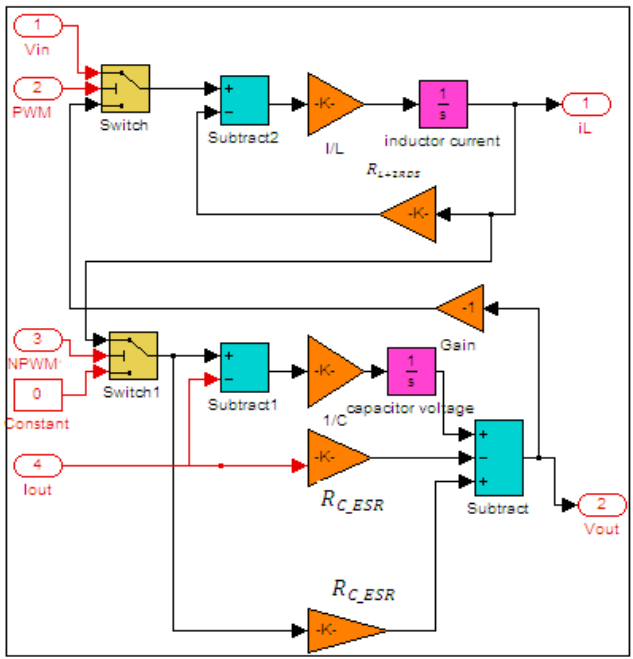

Fig.2 Open-loop of Buck-Boost Converter model

## B. Close-loop synchronous buck-boost Conveter model

### 1) Analog controller

The Figuire.3 is presented the model by SIMULINK/MATLAB of buck-boost converter with analog controller, it uses the compensator of type III-A, and the model aims to regulate the output voltage in 3.24 (V) with variation of input voltage and load.

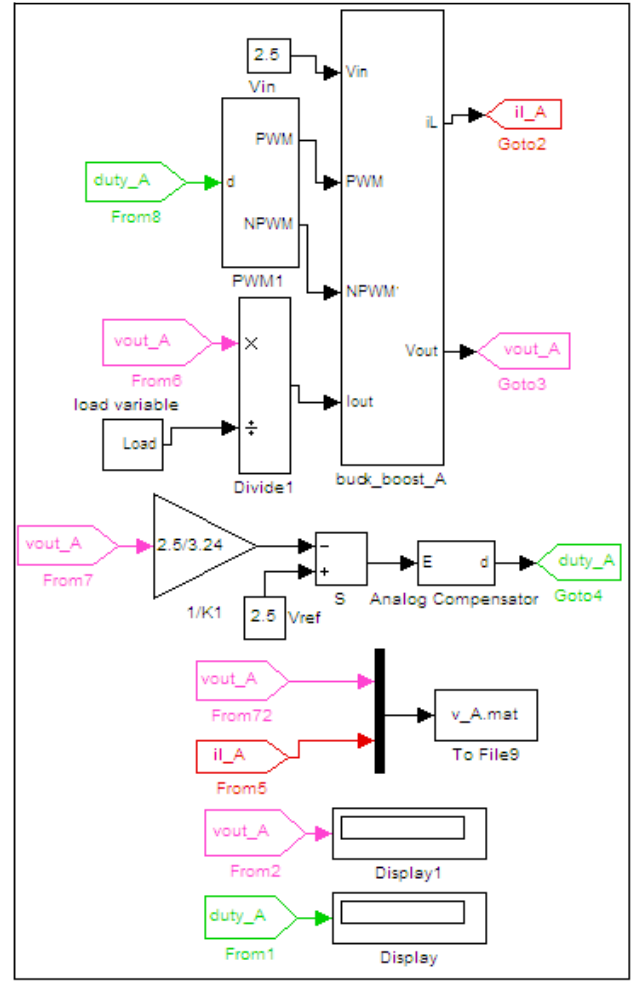

Fig.3 Close- loop buck-boost Converter model Analog controller

### 2) Digital controller

The figure 4 is presented the model by SIMULINK/MATLAB of buck-boost converter with digital controller, the model aims to regulate the output voltage in 3.24 (V), with variation of input voltage and load.

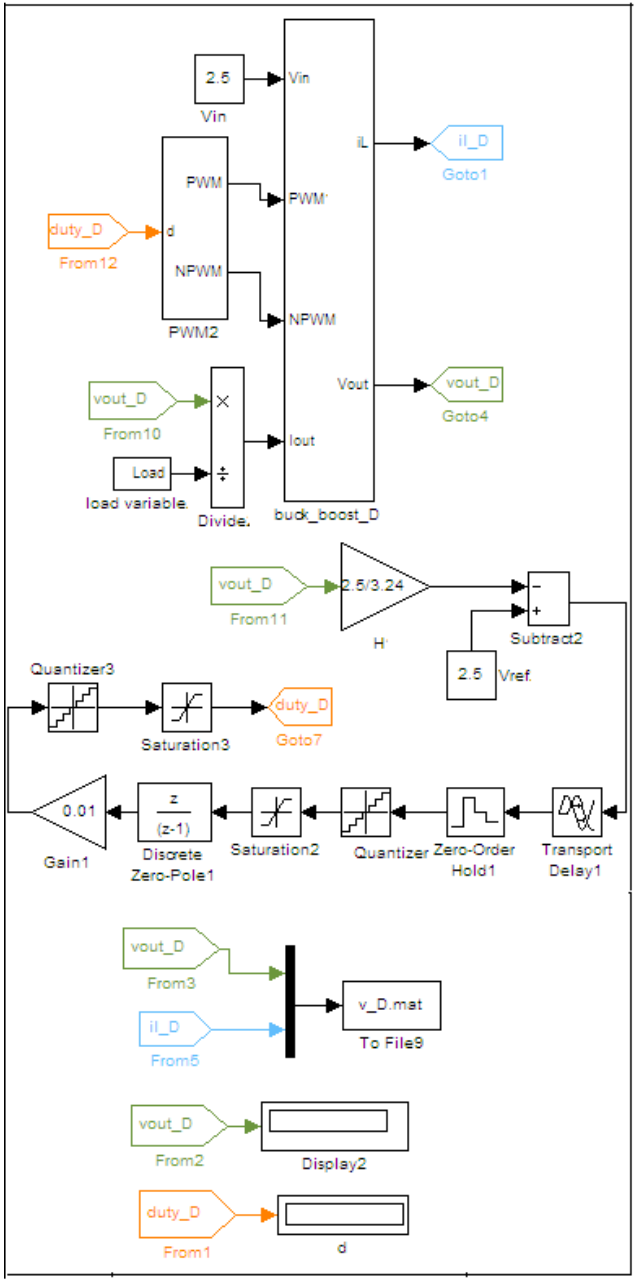

Fig.4 Close-loop buck-boost Converter model digital controller

## III. RUSULTS AND DISCUSSION

In this section, simulation results for Analog controller, digital controller and without feedback on buck-boost converter circuit.

### A. Boost (setup) mode

Table I shows the parameters of buck-boost converter on mode boost of three models, and fig.5 and fig.6 show the inductor current and output voltage waveforms of three models on mode boost.

TABLE I. BUCK-BOOST CONVERTER PARAMETERS ON MODE BOOST

|  | Values without Feedback | Values with Analog controller | Values with Digital Controller |
|---|---|---|---|
| $V_{in}(V)$ | 2.5 | 2.5 | 2.5 |
| $V_{out}$ (V) | 3.24 | 3.24 | 3.4 |
| L (H) | 280e-9 | 1e-6 | 280e-9 |
| C (F) | 250e-9 | 22e-6 | 250e-9 |
| $R_{L+2RDS}(\Omega)$ | 0.5 | 8e-2 | 0.5 |
| $R_{load}(\Omega)$ | 10 | 10 | 10 |
| $R_{C\_ESR}(\Omega)$ | 1e-4 | 60e-3 | 1e-4 |
| Duty cycle | D>0.5 | D>0.5 | D>0.5 |

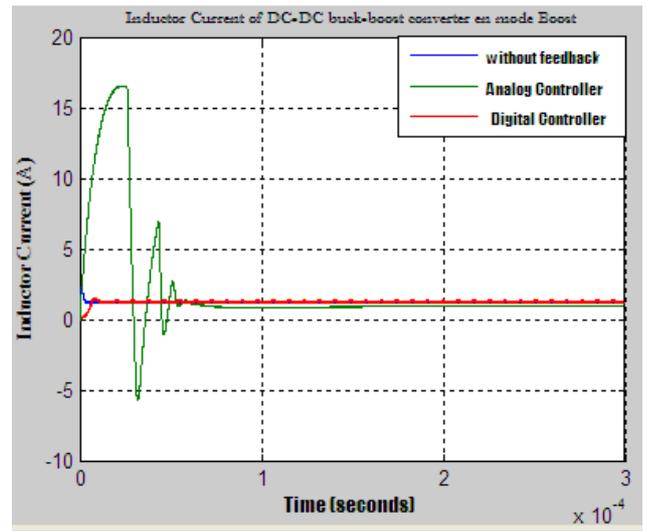

Fig.5 inductor current of buck-boost converter on mode boost

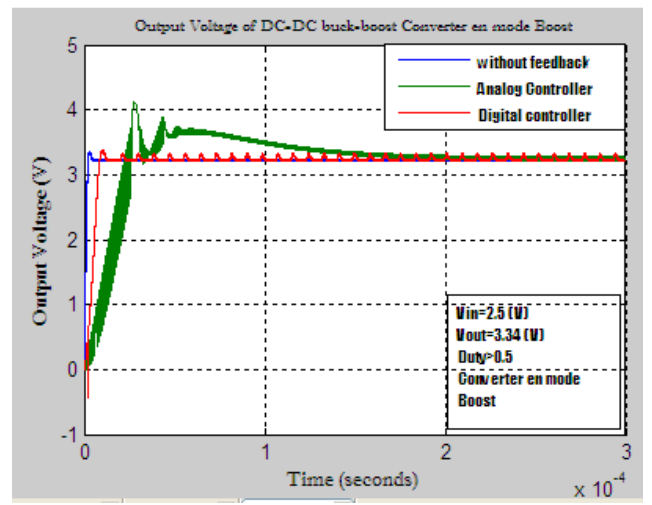

Fig.6 Output voltage of buck-boost converter on mode boost

## B. Buck (step down) mode

Table II shows the parameters of buck-boost converter on mode buck of three models, and fig.7 and fig.8 show the inductor current and output voltage waveforms of three models on mode buck.

TABLE II. BUCK-BOOST CONVERTER PARAMERTER ON MODE BUCK

|  | Values without Feedback | Values with Analog controller | Values with Digital Controller |
|---|---|---|---|
| $V_{in}(V)$ | 5 | 5 | 5 |
| $V_{out}(V)$ | 3.24 | 3.24 | 3.4 |
| L (H) | 280e-9 | 1e-6 | 280e-9 |
| C (F) | 250e-9 | 22e-6 | 250e-9 |
| $R_{L+2RDS}(\Omega)$ | 0.5 | 8e-2 | 0.5 |
| $R_{load}(\Omega)$ | 10 | 10 | 10 |
| $R_{C\_ESR}(\Omega)$ | 1e-4 | 60e-3 | 1e-4 |
| Duty cycle | D<0.5 | D<0.5 | D<0.5 |

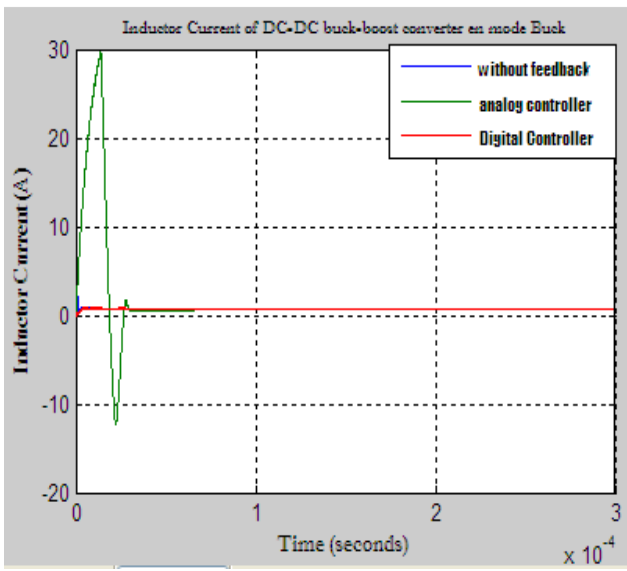

Fig.7 inductor current of buck-boost converter on mode buck

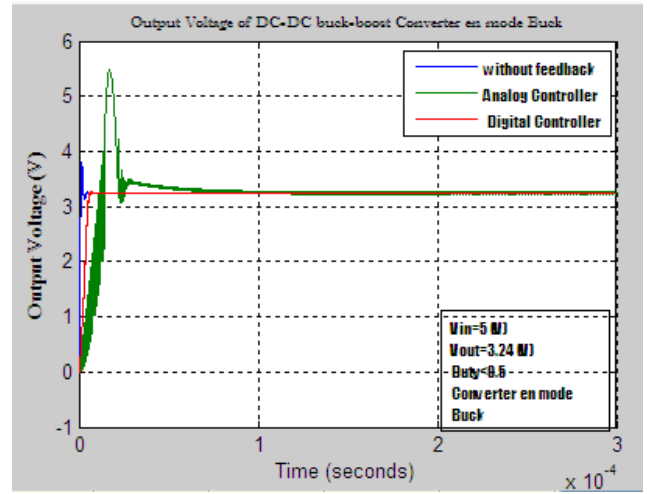

Fig.8 Output voltage of buck-boost converter on mode buck

## C. Comparison between analog controller, digital controller and without feedback.

The circuit has been modeled using Simulink/Matlab, the supply voltage ranges from 2.5V to 5V. the nominal switching frequency is 50 MHz . Experimental results show that the output voltage is regulated in 3.24V of three models and on two mode boost and buck independent of input voltage and load variation.

From figures 5 to 8 show that inductor current and output voltage of three buck-boost converter models in two mode boost and buck, Note that the transit waveforms of digital controller and without feedback are almost the same except for a slight difference, and the waveforms of analog controller model are deferent of previously waveform models.

From tables I and II we note that the values of the components digital controller and without feedback models are equal but they are less than the values component of analog controller model, and the duty cycle are almost equals in three models, in mode boost D>0.5 and in mode buck D<0.5 .

## IV. CONCLUSION

Matlab/Simulink provides an effective environment for modeling and simulation of DC/DC converters. As conclusion digital controller model gives very good dynamic respond compare with analog controller model in two mode buck and boost. And the digital controller model achieves our goal; it minimizes the values of components and conserves same results.